\documentclass[a4paper,12pt]{article}
\usepackage{geometry}
\usepackage[cp1251]{inputenc}
\usepackage{epsfig}
\usepackage{graphicx}

\usepackage{amssymb}

\begin{document}

Astronomy Letters, 2013, Vol. 39, pp. 0-00 \hspace{20pt} Printed 11 
March 2013

\vspace{-8pt}

\section*{
\begin{center}
Magnetically Active Stars in Taurus–Auriga: Photometric Variability
and Basic Physical Parameters
\end{center}
}

\begin{center}
\Large K. N. Grankin
\end{center}

\begin{center}
\it{Crimean Astrophysical Observatory, Nauchny, Crimea, 98409 Ukraine}
\end{center}

\begin{center}
konstantin.grankin@rambler.ru
\end{center}

\subsection*{\center Abstract}

\begin{quote}
The paper presents an analysis of homogeneous long-term photometric 
observations of 28 well-known weak-line T Tauri stars (WTTS) and 60 
WTTS candidates detected by the ROSAT observatory in the direction of 
the Taurus–Auriga star-forming region. 22 known WTTS and 39 WTTS 
candidates are shown to exhibit periodic light variations that are 
attributable to the phenomenon of spotted rotational modulation.
The rotation periods of these spotted stars lie within the range from 
0.5 to 10 days. Significant differences between the long-term 
photometric behaviors of known WTTS and WTTS candidates have been found.
We have calculated accurate luminosities, radii, masses, and ages for 
74 stars. About 33\% of the sample of WTTS candidates have ages younger 
than 10 Myr. The mean distance to 24 WTTS candidates with reliable 
estimates of their radii is shown to be $143\pm 26$ pc. This is in 
excellent agreement with the adopted distance to the Taurus–Auriga 
star-forming region.
\end{quote}

\vspace{15pt}

\textbf{Key words:} {\it stars -- variable, properties, rotation, 
pre-main-sequence stars.}

\subsection*{INTRODUCTION}

The all-sky survey with the ROSAT X-ray space
observatory has revealed a large number of magnetically
active late-type stars in the neighborhoods
of star-forming regions. In particular, Wichmann
et al. (1996) identified 76 candidates for T Tauri
stars (TTS) toward the Taurus–Auriga star-forming
region (SFR). Subsequently, Wichmann et al. (2000)
undertook a detailed study of 58 TTS candidates
based on high-resolution echelle spectroscopy and
proper motions. They found that about 60\% of this
sample could be regarded as pre-main-sequence (PMS)
stars, while the remaining stars are probably
zero-age main-sequence (ZAMS) stars.

Bouvier et al. (1997) undertook the first photometric
monitoring of 58 stars from Wichmann’s list to
measure their rotation periods. They found the photometric
periods for 18 stars. Subsequently, Broeg
et al. (2006) investigated 10 spectroscopic binaries
from Wichmann’s list. They confirmed the periods for
four stars and were the first to detect periodicities for
six objects. Concurrently, Xing et al. (2006) reported
the detection of periods for three more stars. Owing
to these works, the number of stars from Wichmann’s
list with known rotation periods increased
to 27. Grankin et al. (2007a) undertook a long-term
photometric study of 39 objects from Wichmann’s list
during several observing seasons and were able to
find periodic light variations for 22 stars for the first
time. This result allowed the number of stars from
Wichmann’s list with known periods to be almost
doubled.

Thus, on the one hand, there are long-term photometric
measurements for 28 young stars in the
Taurus–Auriga SFR with a known evolutionary status
(see Grankin et al. 2008); 22 stars from this
group exhibit periodic light variations (below, they will
be mentioned as well-known WTTS or stars from
Grankin’s list). On the other hand, there are such
data for 60 magnetically active stars with a controversial
evolutionary status (WTTS candidates or stars
from Wichmann’s list); 39 stars from this second
group exhibit periodic light variations (see Grankin
et al. 2007a).

In this paper, we use homogeneous long-term
photometry for 28 well-known WTTS from Grankin’s
list and 60 WTTS candidates from Wichmann’s list
to determine their basic physical parameters and to
improve the data on these magnetically active stars in
the Taurus–Auriga SFR.

\subsection*{OBSERVATIONS}

All our broadband $BVR$ observations for the stars
from Wichmann’s list were obtained at the Maidanak
Astronomical Observatory (longitude $\rm E4^{h}27^{m}47^{s}$;
latitude $+38^{o}41^{\prime}$; altitude 2709~m). For our long-term
monitoring, we selected 62 stars with $V<14.^{m}4$. More than 5000
$BVR$ magnitudes were obtained for these objects in the period between
1994 and 2006. All our observations were performed with three 
telescopes (one 0.48-m and two 0.6-m reflectors) using identical 
single-channel pulse-counting photometers with FEU-79 photomultiplier
tubes. A detailed description of the observing and data reduction 
technique can be found in Grankin et al. (2008).

The results of the Maidanak program of homogeneous
long-term photometry for 62 WTTS candidates
from Wichmann’s list are presented in Table 1.
The columns in Table 1 give: the star number in
Wichmann’s list, the star name, the epoch of observations,
the number of observing seasons ($N_{\rm s}$), the
number of observations in the $V$ band ($N_{\rm obs}$), the
mean brightness level ($\overline{V_m}$) and its seasonal standard
deviation ($\sigma_{V_m}$), the mean photometric amplitude
($\overline{\Delta V}$, averaged over all seasons) and its seasonal
standard deviation ($\sigma_{\Delta V}$), and the mean colors $\overline{B{-}V}$
and $\overline{V{-}R}$. The series of statistical parameters given
in the last six columns was determined and discussed
in detail previously (see Grankin et al. 2007b, 2008).

\subsection*{BASIC PHOTOMETRIC PROPERTIES} 

The stars of our sample exhibit a low variability
level. Analysis of the mean variability amplitude distribution
(histogram) in the $V$ band for 60 WTTS
candidates from Wichmann’s list showed that it is
sharp-pointed, with $\overline{\Delta V}$ in the range from $0.^{m}05$ to
$0.^{m}30$. Indeed, most of the stars in this sample exhibit
small variability amplitudes (between $0.^{m}05$ and
$0.^{m}15$) and only two stars (3\%) have mean amplitudes
in the $V$ band larger than $0.^{m}2$. The maximum of
the distribution lies near $0.^{m}1$. Here and below, we
excluded two stars (N9 and N72) from our analysis,
because their photometric variability was not caused
by rotational modulation. More detailed information
about these two stars will be given in the section
devoted to the results of our periodogram analysis.

\clearpage
\begin{table*}
\caption{Long-term photometry for WTTS candidates from Wichmann’s list 
(Wichmann et al. 1996)}
\centering
\label{meansp} 
\vspace{2mm}
\begin{small}
\begin{tabular}{r|l|c|c|c|c|c|c|c|c|c} \hline \hline 

\rule{0pt}{2pt}&&&&&&&&&&\\

W96 & {Name} & Epoch & $N_{\rm s}$ & $N_{\rm obs}$ & $\overline{V_m}$ & 
$\sigma_{V_m}$ & $\overline{\Delta V}$ & $\sigma_{\Delta V}$ & 
$\overline{B{-}V}$ & $\overline{V{-}R}$ \\ [5pt]
 \hline
  1 & RX J0400.5+1935 &  1995-2005 &  4 & 103 & 10.19 & 0.014 & 0.12 & 0.036 & 0.94 & 0.82 \\
  2 & RX J0403.3+1725 &  1995-2005 &  4 &  75 & 11.70 & 0.012 & 0.12 & 0.049 & 1.10 & 0.95 \\
  3 & RX J0405.1+2632 &  1999-2005 &  6 & 121 & 11.35 & 0.062 & 0.09 & 0.019 & 0.86 & 0.74 \\
  4 & RX J0405.3+2009 &  1995-2005 &  4 &  95 & 10.41 & 0.033 & 0.11 & 0.037 & 0.91 & 0.77 \\
  5 & RX J0405.7+2248 &  1994-2006 &  5 & 109 &  9.36 & 0.038 & 0.05 & 0.008 & 0.63 & 0.58 \\
  6 & RX J0406.7+2018 &  1994-2006 &  5 &  94 &  9.65 & 0.084 & 0.05 & 0.018 & 0.58 & 0.53 \\
  7 & RX J0406.8+2541 &  2004-2006 &  3 &  53 & 11.91 & 0.048 & 0.30 & 0.029 & 1.38 & 1.28 \\
  8 & RX J0407.8+1750 &  1994-2005 &  4 &  61 & 11.46 & 0.091 & 0.07 & 0.027 & 0.86 & 0.80 \\
  9 & RX J0408.2+1956 &  1994-2005 &  3 &  61 & 13.00 & 0.063 & 0.41 & 0.266 & 1.28 & 1.11 \\
 10 & RX J0409.1+2901 &  1994-2005 & 11 & 259 & 10.56 & 0.018 & 0.10 & 0.029 & 0.84 & 0.75 \\
 11 & RX J0409.2+1716 &  2003-2005 &  3 &  32 & 13.30 & 0.009 & 0.09 & 0.025 & 1.40 & 1.43 \\
 12 & RX J0409.8+2446 &  2003-2005 &  3 &  43 & 13.48 & 0.007 & 0.16 & 0.061 & 1.47 & 1.48 \\
 13 & RX J0412.8+1937 &  1995-2005 &  4 &  76 & 12.57 & 0.029 & 0.09 & 0.010 & 1.30 & 1.18 \\
 14 & RX J0412.8+2442 &  1995-2005 &  4 &  74 & 12.00 & 0.020 & 0.08 & 0.023 & 1.07 & 1.01 \\
 15 & RX J0413.0+1612 &  1994-2006 &  5 &  95 & 11.02 & 0.074 & 0.04 & 0.015 & 0.74 & 0.66 \\
 18 & RX J0415.3+2044 &  1994-2005 & 11 & 224 & 10.65 & 0.031 & 0.10 & 0.044 & 0.72 & 0.66 \\
 19 & RX J0415.8+3100 &  1995-2005 &  4 & 102 & 12.36 & 0.057 & 0.10 & 0.023 & 0.89 & 0.82 \\
 23 & RX J0420.3+3123 &  2004-2006 &  3 &  62 & 12.56 & 0.022 & 0.16 & 0.019 & 1.08 & 0.96 \\
 27 & RX J0423.7+1537 &  1994-2005 & 11 & 225 & 11.25 & 0.026 & 0.10 & 0.017 & 0.93 & 0.81 \\
 29 &RX J0424.8+2643B &  1999-2005 &  6 &  89 & 11.34 & 0.049 & 0.10 & 0.021 & 1.32 & 1.20 \\
 28 &RX J0424.8+2643A &  1999-2005 &  6 &  83 & 11.42 & 0.045 & 0.06 & 0.011 & 1.35 & 1.23 \\
 30 & RX J0427.1+1812 &  1999-2005 &  6 & 100 &  9.42 & 0.044 & 0.07 & 0.034 & 0.56 & 0.53 \\
 31 & RX J0430.8+2113 &  1995-2005 &  4 &  97 & 10.34 & 0.007 & 0.11 & 0.023 & 0.72 & 0.65 \\
 32 & RX J0431.3+2150 &  1994-2005 & 11 & 223 & 10.81 & 0.037 & 0.10 & 0.027 & 0.83 & 0.71 \\
 36 & RX J0432.7+1853 &  1994-2005 &  4 &  68 & 10.81 & 0.020 & 0.09 & 0.044 & 0.82 & 0.70 \\
 37 & RX J0432.8+1735 &  2004-2005 &  2 &  28 & 13.71 & 0.046 & 0.08 & 0.023 & 1.65 & 1.59 \\
 38 & RX J0433.5+1916 &  2003-2005 &  3 &  52 & 13.13 & 0.011 & 0.09 & 0.009 & 1.03 & 0.92 \\
 39 & RX J0433.7+1823 &  2003-2005 &  3 &  35 & 12.07 & 0.019 & 0.05 & 0.008 & 1.06 & 0.95 \\
 40 & RX J0435.9+2352 &  2004-2005 &  2 &  33 & 13.41 & 0.029 & 0.15 & 0.017 & 1.55 & 1.58 \\
 41 & RX J0437.2+3108 &  2003-2005 &  3 &  30 & 13.23 & 0.058 & 0.08 & 0.006 & 1.34 & 1.25 \\
 44 & RX J0438.2+2023 &  1994-2006 &  5 &  76 & 12.20 & 0.008 & 0.10 & 0.037 & 1.12 & 0.97 \\
 45 & RX J0438.2+2302 &  2004-2005 &  2 &  31 & 13.83 & 0.011 & 0.09 & 0.007 & 1.48 & 1.40 \\
 46 & RX J0438.4+1543 &  1995-2005 &  4 &  64 & 13.34 & 0.105 & 0.09 & 0.012 & 1.21 & 1.09 \\
 47 & RX J0438.7+1546 &  1994-2005 &  4 &  51 & 10.83 & 0.074 & 0.10 & 0.048 & 0.99 & 0.87 \\
 48 &RX J0439.4+3332A &  1994-2006 &  5 &  85 & 11.48 & 0.030 & 0.11 & 0.050 & 1.19 & 1.06 \\
 49 & RX J0441.4+2715 &  2003-2005 &  3 &  25 & 13.10 & 0.042 & 0.04 & 0.012 & 0.98 & 0.92 \\
 50 & RX J0441.8+2658 &  1994-2006 &  5 &  61 &  9.59 & 0.088 & 0.04 & 0.009 & 0.60 & 0.58 \\
 51 & RX J0443.4+1546 &  1999-2005 &  6 &  68 & 12.88 & 0.037 & 0.08 & 0.040 & 1.03 & 0.95 \\
 52 & RX J0444.3+2017 &  1994-2005 &  4 &  54 & 12.64 & 0.048 & 0.12 & 0.016 & 1.14 & 1.02 \\
 53 & RX J0444.4+1952 &  1994-2005 &  4 &  53 & 12.57 & 0.017 & 0.09 & 0.033 & 1.49 & 1.41 \\
 54 & RX J0444.9+2717 &  1994-2005 &  4 &  49 &  9.59 & 0.115 & 0.06 & 0.013 & 0.91 & 0.82 \\
 55 & RX J0445.8+1556 &  1994-2005 &  4 &  63 &  9.35 & 0.045 & 0.11 & 0.039 & 0.75 & 0.67 \\
 56 & RX J0446.8+2255 &  2003-2005 &  3 &  36 & 12.88 & 0.023 & 0.13 & 0.016 & 1.43 & 1.38 \\
 57 & RX J0447.9+2755 &  1999-2005 &  6 &  86 & 12.41 & 0.041 & 0.10 & 0.020 & 1.12 & 1.02 \\
 58 & RX J0450.0+2230 &  1995-2005 &  4 &  77 & 11.22 & 0.023 & 0.11 & 0.030 & 0.89 & 0.78 \\ 
\hline
\end{tabular}
\end{small}
\end{table*}

\begin{table*}
{\begin{flushleft}Table 1: continued.\end{flushleft}}
\centering
\label{meansp} 
\vspace{1mm}
\begin{small}
\begin{tabular}{r|l|c|c|c|c|c|c|c|c|c} \hline \hline 

\rule{0pt}{2pt}&&&&&&&&&&\\

W96 & {Name} & Epoch & $N_{\rm s}$ & $N_{\rm obs}$ & $\overline{V_m}$ & 
$\sigma_{V_m}$ & $\overline{\Delta V}$ & $\sigma_{\Delta V}$ & 
$\overline{B{-}V}$ & $\overline{V{-}R}$ \\ [5pt]
 \hline
  59 & RX J0451.8+1758 &  2004-2005 &  2 &  22 & 14.02 & 0.028 & 0.14 & 0.062 & 1.56 & 1.59 \\
 60 &RX J0451.9+2849A &  2004-2005 &  2 &  22 & 13.36 & 0.061 & 0.14 & 0.072 & 1.30 & 1.15 \\
 61 &RX J0451.9+2849B &  2004-2005 &  2 &  22 & 14.13 & 0.013 & 0.10 & 0.018 & 1.15 & 1.02 \\
 62 & RX J0452.5+1730 &  1999-2005 &  6 &  83 & 12.06 & 0.023 & 0.06 & 0.025 & 1.09 & 0.95 \\
 63 & RX J0452.8+1621 &  1999-2005 &  6 &  94 & 11.69 & 0.015 & 0.11 & 0.024 & 1.30 & 1.15 \\
 64 & RX J0452.9+1920 &  1999-2005 &  6 &  90 & 12.16 & 0.083 & 0.06 & 0.024 & 1.09 & 0.99 \\
 65 & RX J0453.1+3311 &  2004-2005 &  2 &  24 & 13.80 & 0.052 & 0.10 & 0.051 & 1.07 & 0.93 \\
 66 & RX J0455.2+1826 &  1994-2005 &  4 &  70 &  9.22 & 0.016 & 0.05 & 0.016 & 0.62 & 0.57 \\
 67 & RX J0455.7+1742 &  1999-2005 &  6 & 100 & 11.21 & 0.015 & 0.08 & 0.019 & 1.02 & 0.89 \\
 68 & RX J0456.2+1554 &  2003-2005 &  3 &  39 & 12.67 & 0.024 & 0.11 & 0.049 & 1.31 & 1.17 \\
 70 & RX J0457.0+1600 &  2004-2005 &  2 &  22 & 14.35 & 0.023 & 0.22 & 0.045 & 1.54 & 1.52 \\
 71 & RX J0457.0+1517 &  1994-2006 &  5 &  81 & 10.30 & 0.013 & 0.05 & 0.018 & 0.66 & 0.61 \\
 72 & RX J0457.0+3142 &  1994-2005 & 11 & 261 & 10.63 & 0.044 & 0.14 & 0.065 & 1.62 & 1.35 \\
 73 & RX J0457.2+1524 &  1994-2006 &  5 &  78 & 10.25 & 0.034 & 0.08 & 0.020 & 0.96 & 0.83 \\
 74 & RX J0457.5+2014 &  1994-2005 &  4 &  67 & 11.06 & 0.051 & 0.11 & 0.035 & 0.87 & 0.78 \\
 75 & RX J0458.7+2046 &  1994-2006 & 12 & 210 & 11.92 & 0.026 & 0.09 & 0.026 & 1.24 & 1.08 \\
 76 & RX J0459.7+1430 &  1999-2006 &  7 & 108 & 11.71 & 0.048 & 0.14 & 0.074 & 1.07 & 0.96 \\
\hline
\end{tabular}
\end{small}
\end{table*}

The mean brightness level of the stars seems very
stable during long-term observations, with a very
small standard deviation $\sigma_{V_m}$. Most of the stars (85\%)
show $\sigma_{V_m}\leq 0.^{m}06$. The remaining stars exhibit $\sigma_{V_m}$
between $0.^{m}08$ and $0.^{m}12$. Likewise, the changes
in the photometric amplitude from season to season
characterized by $\sigma_{\Delta V}$ are small and do not exceed
$0.^{m}06$.

\subsection*{PERIODOGRAM ANALYSIS}

We used three different methods to find the rotation
periods: the $\chi^2$ technique, the \textquotedblleft 
string-length\textquotedblright\
algorithm, and the Lomb–Scargle periodogram analysis
(for more detailed information, see Grankin
et al. 2008). The existence of spurious periods is
the main problem that we faced when searching for
periodic light variations. As a rule, we observed
each object once in a night over 1–2 months during
each season. Such a distribution of observations
in time gives rise to the so-called spurious periods
(Tanner 1948). Both true and spurious periods
produce exactly equivalent (in the statistical sense)
folded light curves. Two or three magnitude estimates
should be obtained during one night to determine
the true period. Such intense observations were performed
for 42 stars from Wichmann’s list. For each
selected star, we made two or three $BVR$ measurements
during each night. The availability of several
short-term monitorings allowed us to calculate the
rate of change in the brightness of a star (\(\Delta V/\Delta t\))
and to find the true periodicity among the whole set
of spurious periods. In addition, such an observing
technique revealed short periods (P \textless 1 days).

\begin{figure}[h]
\epsfxsize=8cm
\vspace{0.6cm}
\hspace{4.5cm}\epsffile{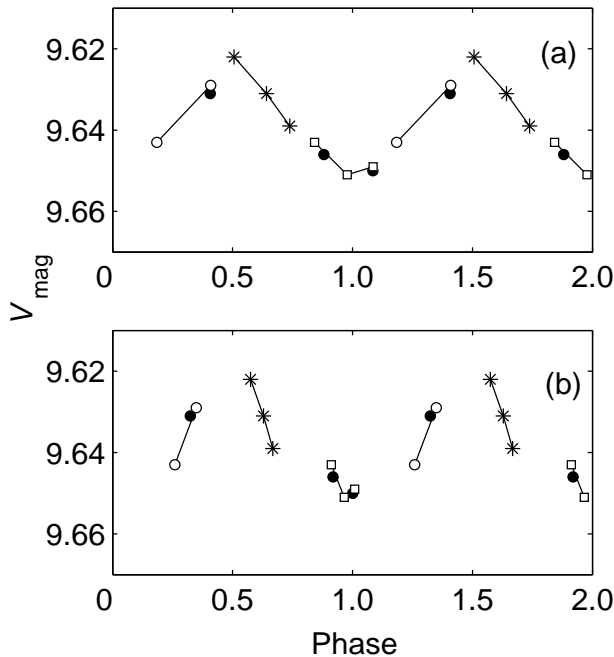}
\caption{\rm \footnotesize {Phase light curve for RX J0441.8+2658 
(number 50 in Wichmann’s list) for two different periods: (a) P = 0.600 
days and (b) P = 1.506 days. The monitoring observations during each 
individual night in 2006 are designated by identical symbols and are 
connected by the solid line. The isolated observations are marked by 
the black circles.}}
\end{figure}

Figure 1 presents an example of the phase light
curves for star N50 from Wichmann’s list. We found
two most probable periods for this star in the 2003 observations:
0.600 (Fig. 1a) and 1.506 days (Fig. 1b).
One of them is true and the other is spurious. Owing
to the monitoring observations performed in 2006, we
were able to show that the shorter period is true.

To reveal the true periods for the stars without
monitoring observations, we used a statistical
method based on Monte Carlo simulations (A. Nemec
and J. Nemec 1985). To guarantee a reliable
significance level, we established a minimum number
of permutations equal to 1000. The periods were
searched for in the interval from 0.5 to 20 days. If
the false alarm probability was between 0.00 and
0.01, then the period was considered true with a 95\%
confidence.

\subsection*{RESULTS OF OUR PERIODOGRAM ANALYSIS}

As a result of the periodogram analysis of our
long-term observations for the stars from Wichmann’s
list, periodic light variations were detected
for the first time in 22 stars and their periods were
improved for 19 more objects. Preliminary data on the
rotation periods of 22 stars can be found in Grankin
et al. (2007a). Table 2 presents information only
about those 19 objects for which the previously published
rotation periods were confirmed or improved.
The columns in Table 2 present the star number
in Wichmann’s list, the star name, the minimum
and maximum amplitudes of periodic light variations
($\Delta V_{\rm min}$ and $\Delta V_{\rm max}$), and the period.

Two stars show periodic light variations that are
not produced by spotted rotational modulation. As is
already well known, the first star, RX J0408.2+1956,
is an eclipsing binary with a period of 3.01 days (see
Bouvier et al. 1997; Broeg et al. 2006). Four observing
seasons allowed the period to be improved. It
is limited by $P = 3.0093\pm 0.0001$ days. The heliocentric
epoch of minimum is $HJD_{\rm min} = 2449658.40$.
The second star, RX J0457.0+3142, is also known
as V501 Aur. Its photometric period is close to
56 days and this is most likely not the rotation period
(see Grankin et al. 2008). We will disregard these
two objects, because we are planning to discuss the
periods attributable only to rotational modulation.

\subsection*{ROTATIONAL MODULATION}

Thirty nine stars from Wichmann’s sample show
periodic light variations during one or more seasons.
To increase our sample of magnetically active stars
with known rotation periods, we used data for 22
WTTS from Grankin et al. (2008). Owing to such a
combination, the sample of stars with known rotation
periods toward the Taurus–Auriga SFR reached 61
objects. Analysis of the distribution (histogram) of
rotation periods for 61 stars from the combined list
showed that the rotation periods lie within the range
from 0.5 to 10 days. In this case, 50 stars ($\sim$82\%)
exhibit periodic variations in the range from 0.5 to
4.5 days and only 11 stars ($\sim$18\%) have rather long
rotation periods in the range from 5 to 10 days. Below,
we will call these two groups of stars rapidly
and slowly rotating ones. The distribution of periods
among the rapidly rotating stars has a noticeable
deficit of objects with periods of about 2 days. This
deficit is noticeable both among the stars from Wichmann’s
list and among the stars from Grankin’s list.

Analysis of our long-term observations revealed
significant differences between the photometric behaviors
of WTTS from Grankin’s and Wichmann’s
lists. Among the 22 spotted WTTS from Grankin’s
list, there are six stars with record amplitudes of
periodic light variations reaching $0.^m4 - 0.^m8$.
in the $V$ band. All these six most active stars show stability of
the phase of minimum light ($\varphi_{\rm min}$) over 5–19 years of
observations (Grankin et al. 2008).
\rm 
\begin{table*}
\caption{WTTS candidates with confirmed or improved periods}
\vspace{6mm}
\centering

\begin{tabular}{r|l|r|r|r|r|r|r|r}
\hline \hline 
\rule{0pt}{2pt}&&&\\
W96 & Name & $ \Delta V_{\rm min}$ &  $ \Delta V_{\rm max}$ &  
\multicolumn{5}{c}{$P$ (days)}\\ \cline{5-9}
    &   &  &      &    this paper &   bo97 &    br06 &    xi06 &  gr08 \\
 \hline
  2 & RX J0403.3+1725 & 0.08 & 0.11 & 0.573  & 0.573 &  0.574 & 0.287 & \\
  7 & RX J0406.8+2541 & 0.28 & 0.33 & 1.6906 &  1.73 &   1.70 & &      \\
  9 & RX J0408.2+1956 & 0.18 & 0.70 & 3.0093 &  3.02 & 3.0109 & &      \\
 10 & RX J0409.1+2901 & 0.06 & 0.16 & 2.662  & 2.74: &        & & 2.662 \\
 11 & RX J0409.2+1716 & 0.07 & 0.09 & 0.613  &       &   0.60 & &      \\
 14 & RX J0412.8+2442 & 0.05 & 0.11 & 0.865  &   6.7 &        & &      \\
 18 & RX J0415.3+2044 & 0.06 & 0.21 & 1.8122 &  1.83 &        & &1.8122 \\
 23 & RX J0420.3+3123 & 0.14 & 0.17 & 4.429  &   4.2 &        & &      \\
 27 & RX J0423.7+1537 & 0.08 & 0.13 & 1.605  & 1.605 &        & & 1.605 \\
 32 & RX J0431.3+2150 & 0.06 & 0.19 & 2.7136 &  2.71 &        & &2.7136 \\
 36 & RX J0432.7+1853 & 0.06 & 0.13 & 1.561  &  1.55 &        & 1.558 & \\
 47 & RX J0438.7+1546 & 0.10 & 0.14 & 3.0789 &  3.07 &        & &      \\
 52 & RX J0444.3+2017 & 0.11 & 0.14 & 1.1320 &  1.15 &        & &      \\
 55 & RX J0445.8+1556 & 0.04 & 0.15 & 1.1040 & 1.104 &        & &      \\
 63 & RX J0452.8+1621 & 0.09 & 0.16 &   3.46 &       &    3.6 & &      \\
 71 & RX J0457.0+1517 & 0.03 & 0.07 & 3.5130 &  3.33 &        & &      \\
 72 & RX J0457.0+3142 & 0.05 & 0.24 &  55.95 & $>$37.6 & $\gg$ 9 & & 55.95    \\
 73 & RX J0457.2+1524 & 0.06 & 0.12 &   1.72 &  2.39 &        & &      \\
 75 & RX J0458.7+2046 & 0.06 & 0.14 & 7.741  &  7.53 &        & & 7.741\\
\hline
\multicolumn{6}{l}{}\\ [-3mm]
\multicolumn{6}{l}{ Note. bo97: Bouvier et al. (1997); br06: Broeg et 
al. (2006);}\\
\multicolumn{6}{l}{xi06: Xing et al. (2006); gr08: Grankin et al. (2008).}\\

\end{tabular}  
\end{table*}

We performed a similar analysis for our long-term
observations of 39 spotted stars from Wichmann’s
list and could not find the objects among them that
exhibited stability of $\varphi_{\rm min}$ over 6–11 years or that
showed significant photometric variability amplitudes
as did the known WTTS. The maximum amplitude
of periodic light variations observed in the stars from
Wichmann’s list was no more than $0.^m28$ (star N7).
And only two objects (N63 and N75) show stability of
$\varphi_{\rm min}$ over 3–4 seasons (see Fig. 2).

\begin{figure}[h]
\epsfxsize=16cm
\vspace{0.6cm}
\hspace{1cm}\epsffile{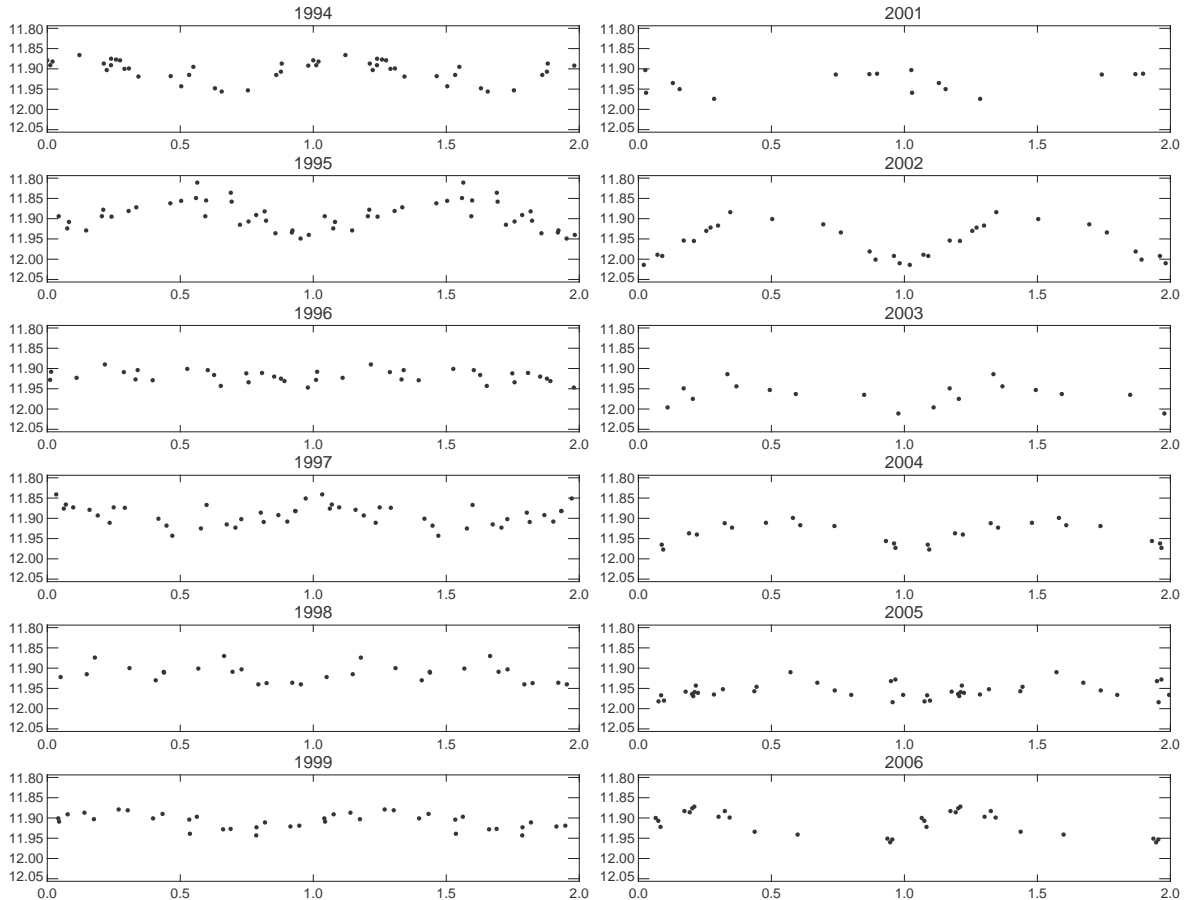}
\caption{\rm \footnotesize {Phase light curves for RX J0458.7+2046 
(number 75 in Wichmann’s list) for each individual season. Periodic light
variations were observed during at least nine seasons. The phase light 
curve retains the phase of minimum light for 4 years (from 2002 to 2005).}}
\end{figure}

There is one more significant difference between
the photometric behaviors of WTTS from Grankin’s
list and Wichmann’s lists. The stars from Wichmann’s
list exhibit periodic light variations not so
often as do WTTS. As a criterion for the frequency of
occurrence of a periodicity, we used a simple parameter
$f =(N_{p}/N_{t})\times100\%$, where $N_p$ is the number
of observing seasons with periodic light variations
and $N_t$ is the total number of observing seasons.
Thus, the most active WTTS mentioned above show
periodic variations in almost every observing season,
i.e., $f=90-100\%$. Other, less active WTTS exhibit
periodic variations with a mean frequency of $68\%$.

In contrast, the mean frequency of occurrence of
periodic variations in the stars from Wichmann’s list
does not exceed 50\%, i.e., periodic variations are
observed in half the observing seasons. Only five
stars showed a fairly frequent occurrence of periodic
variations: N63, N75, N28, N67, and N32, with f =
83, 73, 67, 67, and 64\%, respectively. We hypothesize
that such significant differences in the photometric
behavior of WTTS from Grankin’s and Wichmann’s
lists stem from the fact that these two subgroups
of stars show different activity levels and/or have
slightly different ages.

\subsection*{DETERMINATION OF STELLAR PROPERTIES}

To place the object under study on the Hertzsprung–
Russell (HR) diagram and to estimate its mass
and age, its effective temperature and bolometric
luminosity should be determined with the maximum
possible accuracy.

\subsection*{\it Effective Temperature}

We estimated $T_{\rm eff}$ from the spectral types reported
by Wichmann et al. (1996) using the temperature
calibration between the spectral type and $T_{\rm eff}$ for
main-sequence stars from Tokunaga (2000). Before
choosing this calibration, we analyzed several
different temperature scales for main-sequence
stars from Bessell (1991), Cohen and Kuhi (1979),
de Jager and Nieuwenhuijzen (1987), Kenyon and
Hartmann (1995), and Tokunaga (2000). The general
form of these scales is shown in Fig. 3, where the
$T_{\rm eff}$ -- spectral type diagram is presented. A detailed
description of these scales and estimates of their
intrinsic errors can be found in Ammler et al. (2005).
Analysis showed that the systematic shifts between
different temperature scales could reach 160-210 K and
220-260 K for spectral types G4–G9 and M4–M7,
respectively. In this case, the intrinsic errors of the
temperature scales are comparable to these systematic
shifts (see Table 2 in Ammler et al. (2005)).

\begin{figure}[h]
\epsfxsize=10cm
\vspace{0.6cm}
\hspace{3cm}\epsffile{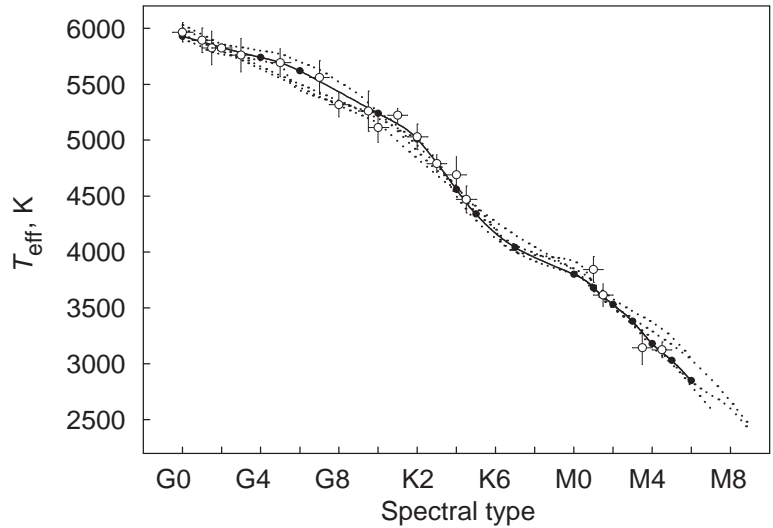}
\caption{\rm \footnotesize {General form of the temperature scales on 
the spectral type versus effective temperature diagram. The calibration 
from Tokunaga (2000) is indicated by the black circles connected by the 
solid line. The remaining calibrations are indicated by the dotted 
lines. The white circles with error bars represent the averaged 
experimental data from Torres et al. (2010).}}
\end{figure}

To choose the most acceptable temperature scale
from these five scales, we used experimental data
on the temperatures and spectral types of 43 G0–
M4.5 dwarfs from Torres et al. (2010). All these
stars are the components of noninteracting eclipsing
binaries for which the mass and radius are known
with an accuracy of $\pm3\%$ or better. Torres et al. (2010)
point out that $T_{\rm eff}$ for these stars have typical errors
of $\sim2\%$. We sorted these 43 stars according their
spectral types and calculated the mean $T_{\rm eff}$ for each
individual subtype. In Fig. 3, these mean $T_{\rm eff}$ are
designated by the white circles. A further analysis
showed that the mean $T_{\rm eff}$ calculated from the experimental
data from Torres et al. (2010) agree with the
temperature calibration from Tokunaga (2000) much
better than with the remaining calibrations. Indeed,
the scatter of mean $T_{\rm eff}$ relative to the calibration from
Tokunaga (2000) is minimal, with an error of $\pm85$~K.
In Fig. 3, the calibration from Tokunaga (2000) is
designated by the black circles connected by the solid
line.

Wichmann et al. (2000) estimate the accuracy of
the spectral classification for bright stars to be about
$\pm0.8$ subtype for G and K stars and about $\pm0.5$ subtype
for M stars. If we use the temperature calibration
from Tokunaga (2000) and take the uncertainty in
the spectral classification for faint stars to be $ \pm1$
subtype, then the corresponding uncertainty in $T_{\rm eff}$
will be $\pm50$~K for G1–G6 stars, $\pm100$~K for G7–K1,
$\pm195$~K for K2–K6, $\pm90$~K for K7–M0, and $\pm160$~K
for M1–M6.

\subsection*{\it Bolometric Luminosity}

As a rule, the stellar bolometric luminosity ($L_{\rm bol}$) is
very difficult to determine, because T Tauri stars often
have optical and near-infrared excesses attributable
to the presence of accretion disks. In our case, these
problems do not arise, because the objects of our
sample show neither infrared excesses nor optical
excess emission. This implies that we can determine
$L_{\rm bol}$ by assuming the entire flux in the $V$ and $R$ bands
to come from the stellar photosphere and calculate
the extinction $A_{V}$ from the color excess as 
$E_{V-R}=(V-R)-(V-R)_{\rm o}$, where $(V-R)$ is the color of
the observed star and $(V-R)_{\rm o}$ is the color of a standard
star of the corresponding spectral type (Kenyon
and Hartmann 1995). Using the standard extinction
law (Johnson 1968), we obtained $A_V = 3.7E_{V-R}$ for
Johnson’s V and R bands.

Finally, $L_{\rm bol}$ was calculated using the well-known
formula 
$\log (L_*/L_\odot) = -0.4(V_{\rm max} -A_V +BC+5 -5\log r-4.72)$, 
where $BC$ is the bolometric correction
from Hartigan et al. (1994) and $r$ is the mean distance
to the Taurus–Auriga SFR (140 pc).

\subsection*{\it Sources of Errors in the Luminosity Estimate}

There are several sources of significant errors in
the $L_{\rm bol}$ estimates: inaccurate photometry, physical
photometric variability, inaccurate spectral classification
and $T_{\rm eff}$ estimation, errors in the adopted colors
for main-sequence dwarfs, the presence of an unresolved
component in the binary system, the lack of
reliable information about the distance to the object
under study, and others. A detailed discussion of the
possible errors and the degree of their influence on the
determination of physical parameters for stars can be
found in Hartigan et al. (1994).

To estimate the magnitude $V_{\rm max}$ and the $V-R$ color, we 
used highly accurate homogeneous
long-term photometry data obtained at the Maidanak
Astronomical Observatory, which is known for
its excellent astroclimate (see, e.g., Ehgamberdiev
et al. 2000). Therefore, $V_{\rm max}$ and $V-R$ were determined
with a high accuracy, at least a few thousandths
of a magnitude. Owing to such a photometric
accuracy, we detected periodic light variations with
amplitudes of the order of $0.^m03-0.^m05$ in the $V$ band,
for example, for RX J0441.8+2658 (see Fig. 1). Thus,
the photometric error makes a minor contribution to
the error in $L_{\rm bol}$.

The photometric variability of the stars from our
sample is attributable primarily to the existence of
extended cool spots. As has been shown above, the
stars of our sample exhibit a small variability amplitude
between $0.^m05$ and $0.^m15$. To minimize the influence
of photometric variability, we used the maximum
brightness ($V_{\rm max}$) and the corresponding color ($V-R$). 
Indeed, at the time of maximum light the visible
stellar surface is least covered with spots. Therefore,
its brightness and color most closely correspond to a
pure photosphere. Thus, we minimized the influence
of photometric variability on the $L_{\rm bol}$ estimate.

Considerably more serious errors can be caused
by the possible existence of an unresolved secondary
component and the uncertainty in the adopted distance.
Thus, the uncertainty due to unresolved binary
systems can imply that we overestimate $L_{\rm bol}$ by a
factor of 2 in the worst case. A no less serious problem
is related to the uncertainty in the distance. Since
the evolutionary status of the stars from our sample
is still unclear and is being actively discussed, their
membership in the Taurus–Auriga SFR has not been
proven. Therefore, we cannot assert with confidence
that all stars of our sample are at a mean distance
of 140 pc. Thus, according to Hipparcos data, the
distance to RX J0406.7+2018 (star N6) is 156 pc,
while the distance to RXJ0441.8+2658 (star N50)
is only 115 pc. This implies that, depending on the
adopted distance (140 or 156 pc in the former case
and 140 or 115 pc in the latter case), the relative error
in $L_{\rm bol}$ can reach tens of percent. The uncertainty in
the distances to these two stars can lead to an error in
$\log L_{\rm bol}$ of the order of $\pm0.1-0.17$ dex.

\begin{figure}[h]
\epsfxsize=12cm
\vspace{0.6cm}
\hspace{2cm}\epsffile{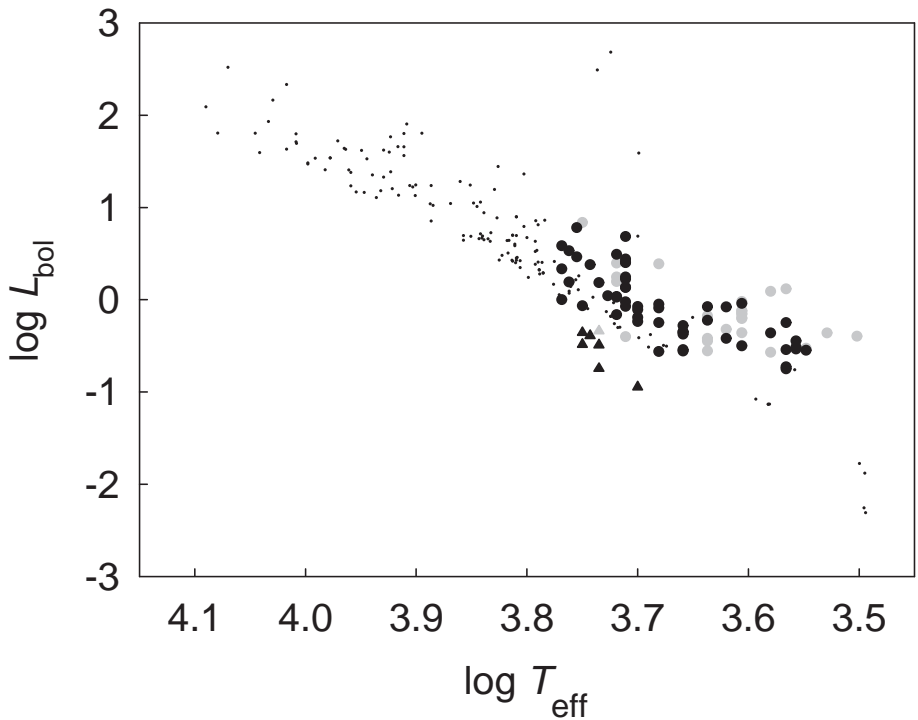}
\caption{\rm \footnotesize {Effective temperature–luminosity diagram. 
About 190 objects from Torres et al. (2010) are marked by the black dots.
The filled black and gray circles represent the stars from Wichmann’s 
and Grankin’s lists, respectively. The objects with underestimated 
luminosities are designated by the black and gray triangles, respectively.}}
\end{figure}

To reveal the stars with definitely underestimated
$L_{\rm bol}$, we used the $T_{\rm eff}$  -- $L_{\rm bol}$ diagram 
presented in Fig. 4. About 190 objects from Torres et al. (2010)
are marked by the black dots. Since the luminosities
of these 190 stars are known with a high accuracy
(at least $\pm0.05$ dex), we compared their positions on
the diagram with those of the stars from Wichmann’s
list (black filled circles) and Grankin’s list (gray filled
circles). Most of the stars from our two lists occupy
the same region on the diagram as do the stars from
the list by Torres et al. (2010). However, there are
seven stars located slightly below the main relation.
We designated them by the black and gray triangles
(N19, N38, N49, N51, N61, N65, and TAP 49). Since
these stars are most likely located appreciably farther
than 140 pc, $L_{\rm bol}$ were underestimated for them.

\subsection*{\it Radius}

The stellar radii were determined by several methods.
First, we estimated the radii ($R_{\rm bol}$) using $T_{\rm eff}$ and
$L_{\rm bol}$.

Second, we used the ratio from Kervella and
Fouqu\'e (2008). These authors gathered the existing
interferometric measurements and broadband
photometry of the nearest dwarfs and subgiants and
obtained polynomial relations between the angular
diameters of these stars and their visible colors. In
particular, they showed that the angular diameter
could be estimated with an accuracy of at least
5\%. We estimated the stellar radii ($R_{\rm KF}$) using the
Kervella–Fouqu\'e calibration by assuming the mean
distance to the Taurus–Auriga SFR to be $\approx$ 140 pc.

\begin{figure}[h]
\epsfxsize=7.8cm
\vspace{0.6cm}
\hspace{4cm}\epsffile{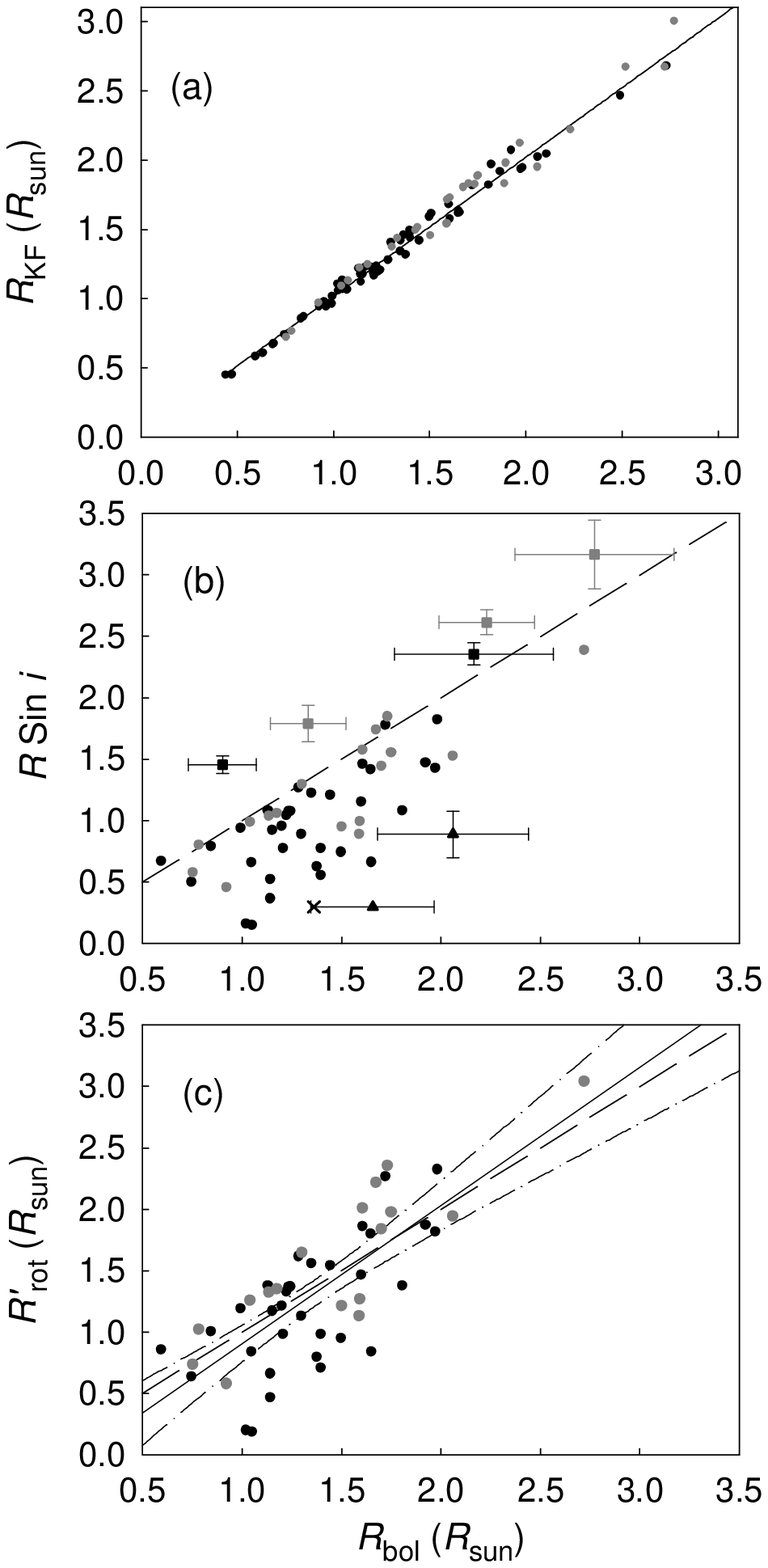}
\caption{\rm \footnotesize {Comparison of the stellar radii $R_{\rm bol}$ 
obtained using $L_{\rm bol}$ and $T_{\rm eff}$ with the radii $R_{\rm KF}$ derived 
from the Kervella–Fouqu\'e relation (a), with the $R \sin i$ estimate 
(b), and with the $R^{'}_{\rm rot}$ estimate (c). The black and gray circles
indicate the stars from Wichmann’s and Grankin’s lists,
respectively. The objects with underestimated and overestimated
$R_{\rm bol}$ are designated by the squares and triangles,
respectively. The cross indicates the position of
object N50 for a distance of 115 pc (for details, see the
text).}}
\end{figure}

The stellar radii determined by these two methods
are compared in Fig. 5a. The estimated radii are in
good agreement with the mean ratio
$\langle R_{\rm KF}/R_{\rm bol}\rangle = 1.004 \pm0.012$.

There is one more possibility for an independent
estimation of the stellar radii without using the distance,
$L_{\rm bol}$, and $T_{\rm eff}$. For the stars with measured
$v\sin i$ and known rotation periods $P_{\rm rot}$, we can estimate
the radii as $ R \sin i = (P_{\rm rot}\times v\sin i) / 2\pi$. In
Fig. 5b, $R_{\rm bol}$ are compared with $ R \sin i$. Since 
$\sin i \leq 1$, the following relation must hold: 
$R \sin i \leq R_{\rm bol}$. The dashed line corresponds to the 
condition $R \sin i = R_{\rm bol}$. It can be seen from the figure that 
this condition holds for most of the stars, with the exception of two
stars from Wichmann’s list (N52 and N55) and three
stars from Grankin’s list (LkCa 3, V410 Tau, and
V836 Tau). They lie well above the dashed line and
are designated by the black and gray squares, respectively.
Since $R \sin i$ are known with greater reliability
than $R_{\rm bol}$, it should be recognized that $R_{\rm bol}$ were
grossly underestimated for these five stars. There are
two possible reasons for this underestimation: either
the distance to these objects is greater than 140 pc or
$ V_{\rm max}$ used to calculate $R_{\rm bol}$ were underestimated. If
these five stars are assumed to be seen face-on, i.e.,
$ \sin i = 1$, then the minimum admissible distance to
these objects cannot be less than 225, 161, 160, 165,
and 168 pc, respectively.

However, this assumption is unacceptable for
V410 Tau, because the distance to this object is
known with a good accuracy and does not exceed
$137\pm17$ pc (Bertout and Genova 2006). The discrepancy
between $R_{\rm bol}$ and $R \sin i$ suggests that even
when we observe the star in its brightest state, it still
has significant spots on the visible surface. In other
words, the true \textquotedblleft unspotted\textquotedblright\ 
magnitude of this star can be brighter than $V_{\rm max}=10.^m57$. 
In particular, Grankin (1999) calculated the absolute unspotted
magnitude of V410 Tau, which is brighter than the
recorded $V_{\rm max}$ by $0.^m33$. In this case, $R_{\rm bol}\simeq2.6R_\odot$,
which corresponds to $R \sin i$ at a mean distance of
about 140 pc. A similar explanation of the discrepancy
between $R_{\rm bol}$ and $R \sin i$ can also be valid
for V836 Tau, whose photosphere is heavily spotted
(Grankin 1998).

There are two other stars (N28 and N50) for which
$R_{\rm bol}$ is considerably larger than $R \sin i$. They are located
in the lower part of the figure and are designated
by the black triangles. It may well be that $L_{\rm bol}$ and
$R_{\rm bol}$ were overestimated for these two objects because
of the distance overestimation or due to the existence
of unresolved components. According to Hipparcos
data, the distance to star N50 is $115\pm18$ pc. If
we estimate $R_{\rm bol}$ for this distance, then the star will
be displaced on the plot leftward, closer to the main
group of stars (cross). It may well be that the second
star (N28) also lies at a closer distance. On the other
hand, Kohler and Leinert (1998) reported that star
N28 consists of four components, while star N50 is a
triple system. Therefore, $R_{\rm bol}$ could be overestimated,
because the contribution from several fairly bright
components was ignored.

For a randomly oriented sample of stars, the mean
$\langle\sin i\rangle\ = \pi/4$ and the stellar radius can be estimated
as $R^{'}_{\rm rot} = R \sin i / \langle\sin i\rangle\ = 
(2P_{\rm rot}\times v \sin i)/\pi^2$. 
$R_{\rm bol}$ and $R^{'}_{\rm rot}$ are compared in Fig. 5c. The solid line
indicates a linear fit to the data with the mean ratio
$\langle R^{'}_{\rm rot}/R_{\rm bol}\rangle\ =  1.125 \pm 0.096$. The 
dash–dotted lines indicate the 95\% confidence region for this linear relation.
The dashed line corresponds to the condition
$R \sin i = R_{\rm bol}$. We excluded the seven stars discussed
above from our regression analysis. It can be seen
from the figure that the regression relation is steeper
than the dashed line that corresponds to the condition
$R^{'}_{\rm rot}=R_{\rm bol}$. This difference can result from a systematic
underestimation of $R^{'}_{\rm rot}$ for stars with small
inclinations, when $\sin i \ll \pi/4$.

To reveal the stars with underestimated $R^{'}_{\rm rot}$, we
used the $T_{\rm eff}$ -- $R^{'}_{\rm rot}$ diagram presented in 
Fig.~6. About 190 objects from Torres et al. (2010) are marked
by the black dots. Since the radii of these objects
are known with a high accuracy (at least $\pm3\%$), we
compared their positions on the diagram with those
of the stars from our sample. Most of the stars from
our sample occupy the same region on the diagram
as do the stars from the list by Torres et al. (2010).
They are designated by the black and gray circles.
However, there are eight stars from Wichmann’s list
and two stars from Grankin’s list that are located
well below the main relation. They are marked by
the black and gray triangles. Since these stars most
likely have small inclinations ($\sin i \ll \pi/4$), 
$R^{'}_{\rm rot}$ were clearly underestimated for them.

\begin{figure}[h]
\epsfxsize=12.0cm
\vspace{0.6cm}
\hspace{2cm}\epsffile{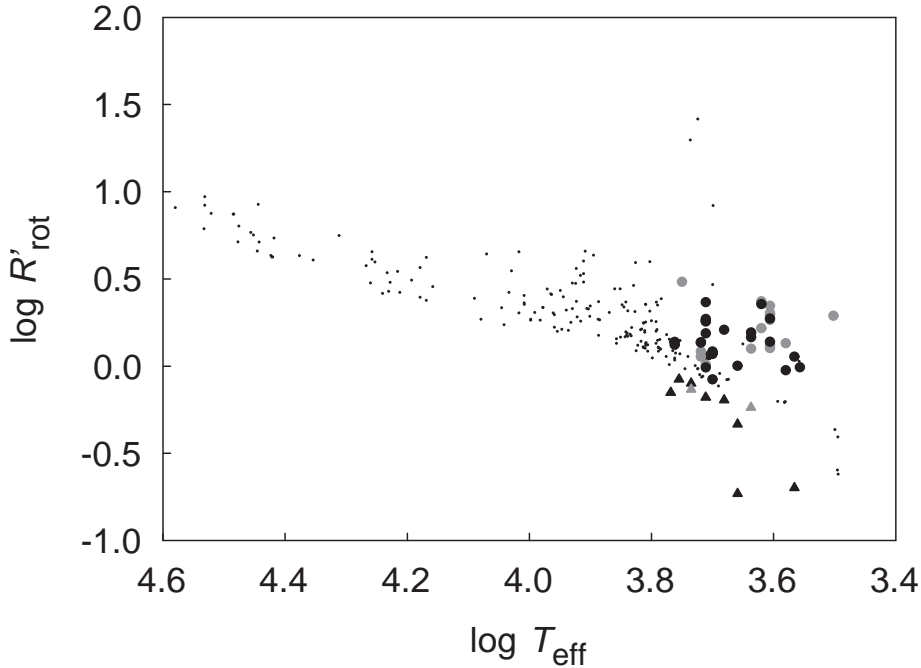}
\caption{\rm \footnotesize {$T_{\rm eff}$ -- $R^{'}_{\rm rot}$ diagram. 
About 190 objects from Torres et al. (2010) are marked by the black 
dots. The stars with reliable $R^{'}_{\rm rot}$ are designated by the 
black and gray circles, while those with underestimated values are 
designated by the black and gray triangles.}}
\end{figure}

Thus, we were able to identify 24 objects with
reliable $R^{'}_{\rm rot}$ estimates and known rotation periods
among the stars from Wichmann’s list. Note that
these estimates do not depend on the adopted distance.
Since the distance plays a key role in determining
$L_{\rm bol}$, we attempted to estimate the mean distance
to these 24 stars using reliable $R^{'}_{\rm rot}$ estimates.
Initially, we calculated the luminosities of these stars
using $R^{'}_{\rm rot}$, $T_{\rm eff}$, and the Stefan–Boltzmann law. Subsequently,
we estimated the distance to each star
using $V_{\rm max}$ and $A_V$ . Finally, we calculated the mean
distance to these 24 stars using the individual distance
estimates. The mean distance to these stars
turned out to be $143\pm26$ pc and to be in excellent
agreement with the adopted distance to the Taurus–Auriga SFR. 
Consequently, we can assert that using the distance $r = 140$ pc is 
quite justified when calculating the basic physical parameters of the 
stars from Wichmann’s list.

\vspace{-2pt}
\subsection*{\it Hertzsprung–Russell Diagram}

To estimate the masses and ages of the magnetically
active stars, we used the grid of evolutionary
tracks for pre-main-sequence stars computed by
Siess et al. (2000). The Hertzsprung–Russell (HR)
diagrams are presented in Fig. 7. For greater clarity,
we show two separate HR diagrams: one for the
stars with reliable rotation periods ($P_{\rm rot}$) and $v \sin i$
(Fig. 7a) and the other for the stars without reliable
$P_{\rm rot}$ or $v \sin i$ (Fig. 7b). For comparison, the diagram
also shows the positions of known WTTS from
Grankin’s list (gray color). The errors in the mass
and age depend on the uncertainties in $T_{\rm eff}$ and $L_{\rm bol}$
adopted in this study and on the object’s position on
the HR diagram. The error in the mass is $\pm0.1\,M_\odot$ and 
$\pm0.2\,M_\odot$ for the objects on convective and radiative
tracks, respectively. The error in the age is of the
order of $\pm1-4$~Myr.

\begin{figure}[h]
\epsfxsize=10.0cm
\vspace{0.6cm}
\hspace{3cm}\epsffile{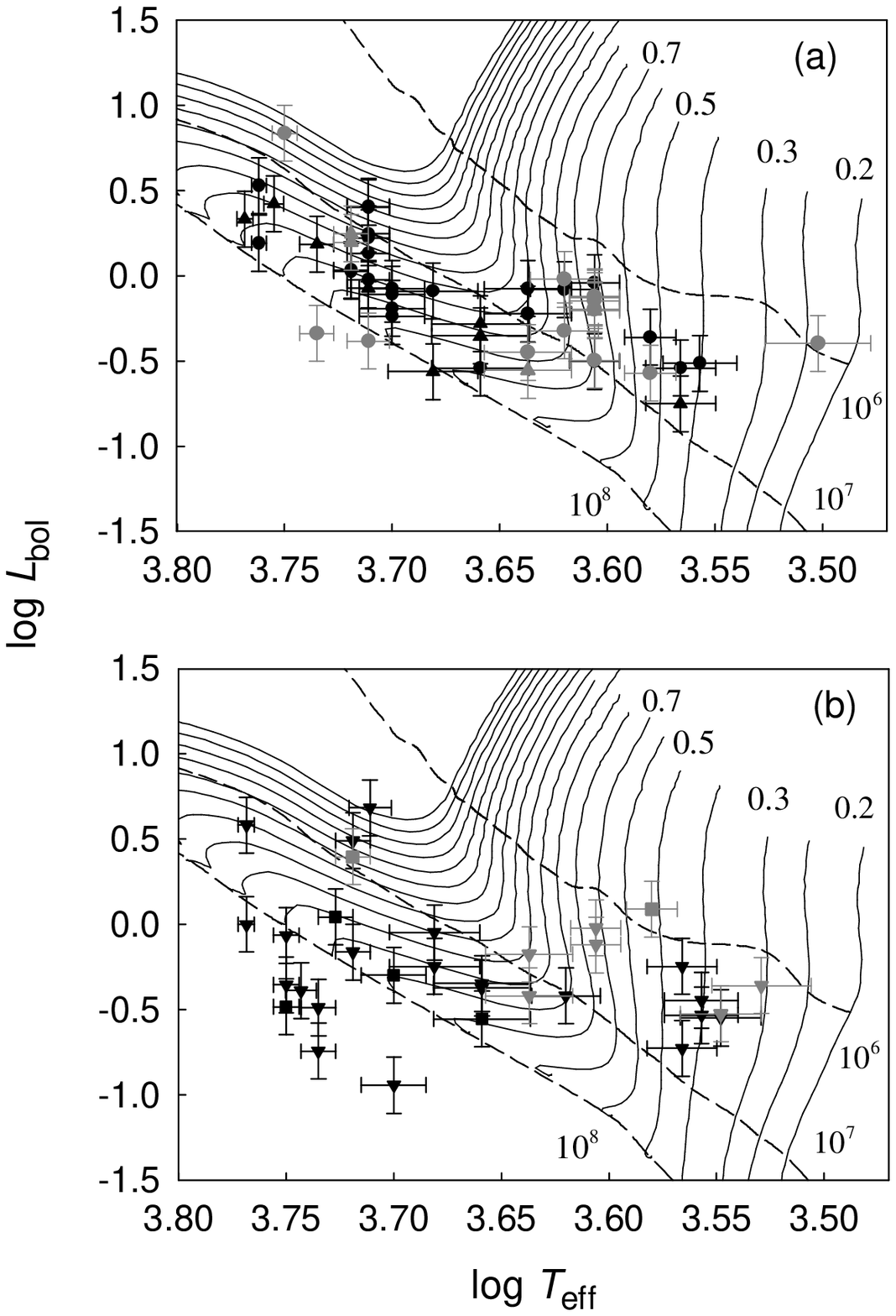}
\caption{\rm \footnotesize {HR diagram for the stars with reliable 
$P_{\rm rot}$ and $v \sin i$ (a) and for the remaining stars (b). The 
black and gray circles represent all stars with reliable $P_{\rm rot}$ 
and good $R^{'}_{\rm rot}$ estimates; the black and gray triangles 
represent the stars with reliable $P_{\rm rot}$ and underestimated 
$R^{'}_{\rm rot}$ ($v \sin i <0.67$). The black and gray squares 
represent the stars with reliable $P_{\rm rot}$ but without $v \sin i$; 
the black and gray overturned triangles represent the stars without 
reliable $P_{\rm rot}$. The errors indicate the $\pm1\sigma$ 
uncertainty for $L_{\rm bol}$ and $T_{\rm eff}$. The solid black lines 
indicate the evolutionary tracks computed with Y = 0.277 and Z = 0.02 
for stars with masses in the range from 0.2 to $2\,M_\odot$ with a mass 
increment of $0.1\,M_\odot$. The dashed lines indicate the isochrones 
for ages of $10^6$, $10^7$, and $10^8$ yr.}}
\end{figure}

The stars from Wichmann’s list have masses in
the range from $0.4\,M_\odot$ to $2.2\,M_\odot$ and ages from 1.5 to
100 Myr. About 33\% of the stars have ages younger
than 10 Myr. Seven stars lie well below the main
sequence on the HR diagram. Their optical spectra
exhibit the $H{\alpha}$ line in absorption and they are G6–G8 dwarfs
(only one star is of spectral type K2). For
these seven stars to be above the main sequence,
their $\log L_{\rm bol}$ must be corrected by more than 0.30–0.55 dex 
and, consequently, they must be located at distances of 
$\backsim 200–280$ pc.

All of the basic parameters for the stars from
Wichmann’s and Grankin’s lists and the intermediate
data used for their calculations are presented in
Tables 3 and 4, respectively. The stars without reliable
data on their luminosities, radii, masses, and ages
are presented in Table 5. The rotation periods were
taken from Grankin et al. (2007a, 2008) and Table 2.
The values of $v \sin i$ were taken from Sartoretti
et al. (1998), Wichmann et al. (2000), Clarke and
Bouvier (2000), Massarotti et al. (2005), and Nguyen
et al. (2009).

\subsection*{CONCLUSIONS}

We analyzed homogeneous long-term photometric
observations of 28 well-known WTTS in
the Taurus–Auriga SFR from Grankin’s list (see
Grankin et al. 2008) and 60 WTTS candidates from
Wichmann’s list (see Wichmann et al. 1996) and
obtained the following results.

Most of the stars from Wichmann’s list (97\%) exhibit
small photometric variability amplitudes in the $V$
band between $0.^{m}05$ and $0.^{m}15$. The mean brightness
level changes from season to season insignificantly
$(\sigma_{V_m}\leq 0.^{m}06)$. Similarly, the changes in the photometric
amplitude ($\sigma_{\Delta V}$) from season to season are
small for most stars (\textless $0.^{m}06$).

We showed that 39 stars from Wichmann’s list
exhibit periodic light variations attributable to the
phenomenon of spotted rotational modulation. We
analyzed the rotation periods for 39 stars from Wichmann’s
list and 22 well-known WTTS from Grankin’s
list. Owing to this combination, the sample of
magnetically active stars with known rotation periods
toward the Taurus–Auriga SFR reached 61 objects.
The rotation periods of the magnetically active stars
lie within the range from 0.5 to 10 days; 50 stars
($\sim$ 82\%) exhibit periodic variations within the range
from 0.5 to 4.5 days and only 11 stars ($\sim 18\%$) have
fairly long rotation periods in the range from 5 to
10 days.

We found significant differences between the
long-term photometric behaviors of 22 WTTS from
Grankin’s list and 39 stars from Wichmann’s list.
There are no very active stars that would exhibit large
amplitudes of periodic light variations and stability
of the phase of minimum light among the objects
from Wichmann’s list. In addition, the stars from
Wichmann’s list exhibit periodic light variations not
so often as do the known WTTS. Thus, the most
active WTTS from Grankin’s list show periodic
variations in almost every observing season, i.e., with
a frequency $f=90-100\%$. Other, less active WTTS
exhibit periodic variations with a mean frequency of
$\sim68\%$. In contrast, the mean detection frequency of
periodic variations in the stars from Wichmann’s list
does not exceed 50\%, i.e., periodic light variations
are observed in no more than half the observing
seasons. We hypothesize that such differences in the
photometric behavior of the known WTTS and stars
from Wichmann’s list are explained by the fact that
these two subgroups of stars show different activity
levels and/or have slightly different ages.

\begin{table*}
\caption{Basic parameters of the WTTS candidates and the intermediate 
data used for their calculations. The mass and age estimates were 
obtained in comparison with the theoretical models and tracks computed 
by Siess et al. (2000)}
\centering
\label{meansp} 
\vspace{5mm}
\begin{scriptsize}
\begin{tabular}{r|l|c|c|c|c|c|c|l|c|c|c|c|r} \hline \hline 

\rule{0pt}{2pt}&&&&&&&&&&&&&\\

W96 & $V_{\rm max}$ & $V-R$ & Sp.  & $\rm E_{V-R}$ & $A_V$ & B.C. & 
$v\sin i$ & $P_{\rm rot}$ & $T_{\rm eff}$ & $ L_{\rm bol}$ & $R$ & 
$M$ & $t$, $10^6$ \\ [1pt]
 & & &type & & & & {($\rm km~c^{-1}$)} & ({days}) & (K) & ($L_\odot$) & 
 ($R_\odot$) & ($M_\odot$) &  (yr) \\
 \hline
 1& 10.144 & 0.804 & K1 &   0.12 & 0.46 & -0.25 &  79 & 1.1683 & 5145 & 2.55 & 1.98 & 1.60 & 7.24   \\
 2 & 11.616 & 0.943 & K3 &   0.14 & 0.53 & -0.41 & 112 & 0.573  & 4801 & 0.81 & 1.28 & 1.15 & 14.60  \\
 3 & 11.229 & 0.740 & K2 &   0.00 & 0.00 & -0.30 &  17 & 1.961  & 5010 & 0.64 & 1.05 & 0.98 & 28.40  \\
 4 & 10.327 & 0.745 & K1 &   0.06 & 0.24 & -0.25 &  25 & 2.86   & 5145 & 1.76 & 1.65 & 1.40 & 10.50  \\
 5 &  9.310 & 0.558 & G3 &   0.03 & 0.10 & -0.08 &  67 & 0.816  & 5786 & 3.39 & 1.81 & 1.38 & 14.40  \\
 6 &  9.667 & 0.517 & G1 &   0.00 & 0.00 & -0.06 &  26 & 1.079  & 5876 & 2.15 & 1.40 & 1.22 & 21.20  \\
 7 & 11.834 & 1.256 & K7 &   0.11 & 0.39 & -0.89 &  44 & 1.6906 & 4040 & 0.91 & 1.92 & 0.73 & 1.96   \\
 8 & 11.482 & 0.800 & K3 &   0.00 & 0.00 & -0.41 &  24 &        & 4801 & 0.56 & 1.07 & 1.00 & 23.50  \\
10 & 10.494 & 0.713 & K1 &   0.03 & 0.12 & -0.25 &  23 & 2.662  & 5145 & 1.35 & 1.44 & 1.28 & 14.60  \\
11 & 13.266 & 1.403 & M1 &   0.00 & 0.01 & -1.45 &  73 & 0.616  & 3680 & 0.29 & 1.30 & 0.45 & 3.27   \\
12 & 13.420 & 1.478 & M1.5 & 0.03 & 0.10 & -1.58 &   7 & 5.58   & 3605 & 0.31 & 1.40 & 0.40 & 2.45   \\
13 & 12.526 & 1.142 & K6 &   0.07 & 0.27 & -0.76 &  12 &        & 4166 & 0.38 & 1.16 & 0.91 & 11.00  \\
14 & 11.960 & 1.003 & G9 &   0.39 & 1.46 & -0.16 &  -- & 0.865  & 5337 & 1.10 & 1.21 & 1.09 & 23.40  \\
15 & 10.947 & 0.638 & G1 & 0.12 & 0.44 & -0.06 &   9 &         & 5876 & 1.00 & 0.95 & 1.20 & 100
\\
18 & 10.557 & 0.640 & K0 &   0.00 & 0.00 & -0.19 &  30 & 1.812  & 5240 & 1.08 & 1.24 & 1.12 & 22.40  \\
23 & 12.478 & 0.941 & K4 &   0.04 & 0.15 & -0.52 &   9 & 4.429  & 4560 & 0.29 & 0.85 & 0.83 & 37.95  \\
27 & 11.192 & 0.786 & K2 &   0.05 & 0.17 & -0.30 &  29 & 1.605  & 5010 & 0.78 & 1.15 & 1.07 & 22.70  \\
29 & 11.293 & 1.148 & K0 &   0.51 & 1.88 & -0.19 &  55 &        & 5240 & 3.10 & 2.11 & 1.67 & 7.01   \\
30 &  9.343 & 0.540 & G5 &   0.00 & 0.00 & -0.09 &  48 & 0.6962 & 5687 & 3.01 & 1.76 & 1.35 & 14.60  \\
31 & 10.277 & 0.623 & G8 &   0.04 & 0.16 & -0.13 &  43 & 0.736  & 5438 & 1.53 & 1.38 & 1.20 & 18.70  \\
32 & 10.704 & 0.675 & K0 &   0.04 & 0.13 & -0.19 &  20 & 2.7136 & 5240 & 1.06 & 1.23 & 1.11 & 22.40  \\
36 & 10.761 & 0.680 & K1 &   0.00 & 0.00 & -0.25 &  25 & 1.561  & 5145 & 0.95 & 1.21 & 1.10 & 21.40  \\
37 & 13.712 & 1.548 & M2 &   0.05 & 0.18 & -1.71 &  11 &        & 3530 & 0.28 & 1.40 & 0.37 & 2.38   \\
39 & 12.051 & 0.911 & G6 & 0.36 & 1.34 & -0.10 &  -- &         & 5620
& 0.86 & 0.96 & 1.10 & 100\\
40 & 13.374 & 1.511 & M1.5 & 0.06 & 0.23 & -1.58 &   6 &        & 3605 & 0.36 & 1.51 & 0.40 & 2.02   \\
41 & 13.135 & 1.232 & K4 &   0.33 & 1.23 & -0.52 &  10 &        & 4560 & 0.42 & 1.03 & 0.95 & 23.00  \\
44 & 12.141 & 0.955 & K2 &   0.22 & 0.80 & -0.30 &  16 & 2.96   & 5010 & 0.58 & 0.99 & 0.95 & 32.30  \\
45 & 13.788 & 1.418 & M1 &   0.02 & 0.07 & -1.45 &   8 &        & 3680 & 0.19 & 1.05 & 0.44 & 5.66   \\
46 & 13.234 & 1.061 & K3 &   0.26 & 0.97 & -0.41 &  10 & 2.535  & 4801 & 0.27 & 0.75 & 0.84 & >100   \\
47 & 10.792 & 0.855 & K1 &   0.18 & 0.65 & -0.25 &  24 & 3.0789 & 5145 & 1.67 & 1.60 & 1.38 & 11.60  \\
48 & 11.397 & 1.027 & K5 &   0.04 & 0.14 & -0.62 &  24 & 2.425  & 4340 & 0.84 & 1.60 & 1.10 & 4.91   \\
53 & 12.526 & 1.401 & M1 &   0.00 & 0.00 & -1.45 &   7 &        & 3680 & 0.56 & 1.82 & 0.45 & 1.44   \\
54 &  9.490 & 0.815 & K1 &   0.14 & 0.50 & -0.25 &  80 &        & 5145 & 4.83 & 2.73 & 2.05 & 3.83   \\
56 & 12.802 & 1.355 & M0 &   0.08 & 0.28 & -1.17 &  10 & 3.762  & 3800 & 0.44 & 1.50 & 0.53 & 2.49   \\
57 & 12.362 & 0.995 & K0 &   0.36 & 1.31 & -0.19 &  27 &        & 5240 & 0.69 & 0.99 & 0.96 & 35.20  \\
58 & 11.166 & 0.757 & K1 &   0.08 & 0.29 & -0.25 &  55 & 0.4778 & 5145 & 0.85 & 1.14 & 1.05 & 25.10  \\
59 & 13.893 & 1.592 & M1.5 & 0.14 & 0.53 & -1.58 &  17 &        & 3605 & 0.29 & 1.36 & 0.40 & 2.60   \\
60 & 13.281 & 1.148 & K4 &   0.25 & 0.92 & -0.52 &  -- & 0.921  & 4560 & 0.28 & 0.83 & 0.82 & 43.90  \\
62 & 12.000 & 0.940 & K4 &   0.04 & 0.15 & -0.52 &   9 & 0.8204 & 4560 & 0.44 & 1.05 & 0.98 & 22.00  \\
63 & 11.626 & 1.128 & K6 &   0.06 & 0.21 & -0.76 &  26 & 3.46   & 4166 & 0.83 & 1.72 & 0.89 & 3.13   \\
64 & 12.054 & 0.958 & K4 &   0.06 & 0.21 & -0.52 &   6 &        & 4560 & 0.45 & 1.06 & 0.98 & 21.20  \\
66 &  9.192 & 0.558 & G1 &   0.04 & 0.14 & -0.06 &  81 &        & 5876 & 3.83 & 1.87 & 1.40 & 14.20  \\
67 & 11.153 & 0.846 & K3 &   0.05 & 0.17 & -0.41 &  19 &        & 4801 & 0.89 & 1.35 & 1.20 & 13.00  \\
68 & 12.595 & 1.150 & K7 &   0.00 & 0.00 & -0.89 & 9.7 & 5.64   & 4040 & 0.31 & 1.13 & 0.80 & 10.20  \\
70 & 14.234 & 1.524 & M1 &   0.12 & 0.46 & -1.45 &   9 & 0.884  & 3680 & 0.18 & 1.02 & 0.44 & 6.00   \\
71 & 10.293 & 0.595 & G3 &   0.06 & 0.24 & -0.08 &  15 & 3.513  & 5786 & 1.55 & 1.22 & 1.16 & 25.40  \\
73 & 10.200 & 0.816 & K1 &   0.14 & 0.50 & -0.25 &  42 & 1.72   & 5145 & 2.52 & 1.97 & 1.60 & 7.24   \\
74 & 10.995 & 0.756 & K2 &   0.02 & 0.06 & -0.30 &  33 & 1.46   & 5010 & 0.84 & 1.20 & 1.10 & 20.10  \\
75 & 11.841 & 1.047 & K5 &   0.06 & 0.21 & -0.62 &   8 & 7.741  & 4340 & 0.60 & 1.35 & 1.06 & 8.23   \\
76 & 11.685 & 0.903 & K4 &   0.00 & 0.01 & -0.52 &  15 & 1.2308 & 4560 & 0.52 & 1.14 & 1.05 & 16.40  \\

\hline
\end{tabular}
\end{scriptsize}
\end{table*}

\begin{table*}
\caption{Basic parameters of the known WTTS and the intermediate data 
used for their calculations. The mass and age estimates were obtained 
in comparison with the theoretical models and tracks computed by Siess 
et al. (2000)}

\centering
\label{meansp} 
\vspace{5mm}
\begin{scriptsize}
\begin{tabular}{r|l|c|c|c|c|c|c|l|c|c|c|c|r} \hline \hline 

\rule{0pt}{2pt}&&&&&&&&&&&&&\\

Name & $V_{\rm max}$& $V-R$ & Sp.  &$\rm E_{V-R}$ &$A_V$ & B.C. & 
$ v\sin i$ & $P_{\rm rot}$ & $T_{\rm eff}$ & $ L_{\rm bol}$ & $R$ & 
$M$ & $t$, $10^6$ \\ [1pt]
 & & &type & & & & {($\rm km~c^{-1}$)} & ({days}) & (K) & ($L_\odot$) & 
 ($R_\odot$) &  ($M_\odot$) & (yr)
  \\ [1pt]
 \hline 
Anon 1   & 13.374 &  1.814 &  M0 &  0.53 &  1.98 &  -1.17 &      & 6.493   & 3800 &      1.23 &  2.52 &  0.52 &  0.81 \\ 
HD283572&  8.914 &  0.673 &  G6 &  0.12 &  0.46 &  -0.10 & 79   & 1.529   & 5620 &      6.87 &  2.72 &  1.84 &  6.26 \\
LkCa 1   & 13.681 &  1.700 &  M4 &  0.00 &  0.00 &  -2.24 & 30.9 & 2.497   & 3180 &      0.40 &  2.06 &  0.24 &  0.37 \\
LkCa 4   & 12.285 &  1.297 &  K7 &  0.15 &  0.54 &  -0.89 & 26.1 & 3.374   & 4040 &      0.74 &  1.73 &  0.74 &  2.60 \\
LkCa 5   & 13.476 &  1.500 &  M2 &  0.00 &  0.00 &  -1.71 & 37   &         & 3530 &      0.30 &  1.44 &  0.37 &  2.26 \\
LkCa 7   & 12.144 &  1.268 &  K7 &  0.12 &  0.44 &  -0.89 & 12.9 & 5.6638  & 4040 &      0.71 &  1.70 &  0.74 &  2.70 \\
LkCa 14  & 11.641 &  1.025 &  K5 &  0.03 &  0.13 &  -0.62 & 21.9 &         & 4340 &      0.67 &  1.43 &  1.09 &  7.00 \\
LkCa 16  & 12.335 &  1.405 &  K7 &  0.26 &  0.94 &  -0.89 & 6.9  &         & 4040 &      0.96 &  1.97 &  0.72 &  1.81 \\
LkCa 19  & 10.807 &  0.851 &  K0 &  0.21 &  0.78 &  -0.19 & 20.1 & 2.236   & 5240 &      1.76 &  1.59 &  1.34 &  12.80 \\
LkCa 21  & 13.451 &  1.648 &  M3 &  0.05 &  0.18 &  -1.92 & 60   &         & 3380 &      0.44 &  1.89 &  0.31 &  1.40 \\
TAP 4    & 12.164 &  0.804 &  K1 &  0.12 &  0.46 &  -0.25 & 83.9 & 0.482   & 5145 &      0.40 &  0.78 &  0.89 &  73.00 \\
TAP 9    & 12.122 &  0.990 &  K5 &  0.00 &  0.00 &  -0.62 & 36   &         & 4340 &      0.38 &  1.08 &  0.95 &  16.60 \\
TAP 26   & 12.194 &  0.990 &  K5 &  0.00 &  0.00 &  -0.62 & 70   & 0.7135  & 4340 &      0.36 &  1.04 &  0.93 &  18.60 \\
TAP 35   & 10.174 &  0.647 &  K0 &  0.01 &  0.03 &  -0.19 & 17.6 & 2.734   & 5240 &      1.57 &  1.50 &  1.28 &  14.50 \\
TAP 40   & 12.535 &  1.011 &  K5 &  0.02 &  0.08 &  -0.62 & 14.8 & 1.5548  & 4340 &      0.28 &  0.92 &  0.86 &  26.20 \\
TAP 41   & 12.028 &  1.073 &  K6 &  0.00 &  0.01 &  -0.76 & 27   & 2.425   & 4166 &      0.48 &  1.30 &  0.92 &  8.14  \\
TAP 45   & 13.135 &  1.299 &  K7 &  0.15 &  0.55 &  -0.89 & 5.3  & 9.909   & 4040 &      0.32 &  1.14 &  0.79 &  10.00  \\
TAP 50   & 10.091 &  0.759 &  K0 &  0.12 &  0.44 &  -0.19 &      & 3.039   & 5240 &      2.49 &  1.89 &  1.54 &  8.70  \\
TAP 57 & 11.530 &  1.106 &  K6 &  0.04 &  0.13 &  -0.76 & 10   & 9.345   & 4166 &      0.95 &  1.85 &  0.88 &  2.43  \\
V819 Tau & 12.796 &  1.405 &  K7 &  0.26 &  0.94 &  -0.89 & 9.1  & 5.53113 & 4040 &      0.62 &  1.59 &  0.76 &  3.30  \\
V826 Tau & 12.069 &  1.265 &  K7 &  0.12 &  0.43 &  -0.89 & 4.2  &         & 4040 &      0.76 &  1.75 &  0.74 &  2.50  \\
V827 Tau & 12.242 &  1.311 &  K7 &  0.16 &  0.60 &  -0.89 & 20.9 & 3.75837 & 4040 &      0.76 &  1.75 &  0.74 &  2.50  \\
V830 Tau & 11.933 &  1.177 &  K7 &  0.03 &  0.10 &  -0.89 & 29.1 & 2.74101 & 4040 &      0.64 &  1.61 &  0.75 &  3.13  \\
VY Tau   & 13.541 &  1.413 &  M0 &  0.13 &  0.49 &  -1.17 & 10   & 5.36995 & 3800 &      0.27 &  1.18 &  0.54 &  5.10  \\

\hline
\end{tabular}
\end{scriptsize}
\end{table*}

\begin{table*}
\caption{List of stars without reliable data on their luminosities, 
radii, masses, and ages}

\centering
\label{meansp} 
\vspace{5mm}
\begin{scriptsize}
\begin{tabular}{r|l|c|c|c|c|c|c|l|c|l} \hline \hline 

\rule{0pt}{2pt}&&&&&&&&&&\\

W96/name & $V_{\rm max}$& V-R & Sp. & $\rm E_{V-R}$ & $A_V$ & B.C. & 
$ v\sin i$ & $P_{\rm rot}$ & $ T_{\rm eff}$ &  Notes    \\ [1pt]
 & & &type & & & & ({$\rm km~c^{-1}$}) & ({days}) & (K)       \\
 \hline
 9 & 12.859 & 1.107 & K2 & 0.37 & 1.36 & -0.30 &  75 &  3.0093 & 5010 &  eclipsing binary \\           
19 & 12.347 & 0.795 & G6 & 0.25 & 0.91 & -0.10 &  31 &         & 5620 &  below MS, $ \rm r > 210$ pc\\
28 & 11.346 & 1.151 & K1 & 0.47 & 1.74 & -0.25 &  14 &  3.21   & 5145 &  $\rm R_{bol} > R \sin i$  \\
38 & 13.079 & 0.906 & G6 & 0.36 & 1.32 & -0.10 &  58 &  0.5854 & 5620 &  below MS, $\rm r > 240$ pc \\
49 & 13.060 & 0.919 & G8 & 0.34 & 1.26 & -0.13 &  34 &         & 5438 &  below MS, $\rm r > 210$ pc\\
50 &  9.613 & 0.567 & G7 & 0.00 & 0.00 & -0.12 &  25 &  0.6    & 5535 &  $\rm R_{bol} > R \sin i$ \\
51 & 12.831 & 0.917 & G7 & 0.35 & 1.29 & -0.12 &  29 &         & 5535 &  below MS, $\rm r > 200$ pc \\
52 & 12.594 & 1.003 & K1 & 0.32 & 1.20 & -0.25 &  65 &  1.132  & 5145 &  $\rm R_{bol} < R \sin i$,  $\rm r > 225$ pc \\
55 &  9.334 & 0.659 & G5 & 0.12 & 0.44 & -0.09 & 114 &  1.104  & 5687 &  $\rm R_{bol} < R \sin i$, $\rm r > 161$ pc \\
61 & 14.108 & 1.009 & K2 & 0.27 & 1.00 & -0.30 &  47 &         & 5010 &  below MS, $\rm r > 265$ pc \\
65 & 13.714 & 0.923 & G8 & 0.34 & 1.27 & -0.13 &  -- &         & 5438 &  below MS, $\rm r > 280$ pc\\
72 & 10.566 & 1.310 & K2 & 0.57 & 2.11 & -0.30 &  27 &  55.95  & 5010 &  $\rm P > P_{rot}$\\
LkCa 3 & 11.978 & 1.500 & M1 & 0.10 & 0.37 & -1.45 &  22 &  7.35  & 3680 &  $\rm R_{bol} < R \sin i$\\
V410 Tau & 10.567 & 0.983 & K3 & 0.18 & 0.68 & -0.41 &  71 &  1.87197  & 4801 &  $\rm R_{bol} < R \sin i$\\
V836 Tau & 13.060 & 1.372 & K7 & 0.22 & 0.82 & -0.89 &  12 &  6.75791  & 4040 &  $\rm R_{bol} < R \sin i$\\
TAP 49 & 12.575 & 0.890 & G8 & 0.31 & 1.15 & -0.13 &  8.8 &  3.32  & 5438 & below MS, $\rm r > 180$ pc \\

\hline
\end{tabular}
\end{scriptsize}
\end{table*}

For all stars, we estimated $T_{\rm eff}$ using the temperature
calibration from Tokunaga (2000). We showed
that the errors in the effective temperature could reach
$\pm50$~K for G1–G6 stars, $\pm100$~K for G7–K1, $\pm195$~K
for K2–K6, $\pm90$~K for K7–M0, and $\pm160$~K for M1–
M6.

We calculated the stellar luminosity $L_{\rm bol}$ by assuming
all stars to be at the mean distance of
the Taurus–Auriga SFR. We discussed the various
sources of errors in the $L_{\rm bol}$ estimates and attempted
to minimize some of these errors. We showed that the
errors in $\log L_{\rm bol}$ could reach $\pm0.1- 0.17$ dex. Using
the $T_{\rm eff}-L_{\rm bol}$ diagram constructed for 190 stars from
Torres et al. (2010) with well-known $T_{\rm eff}$ and $L_{\rm bol}$,
we were able to reveal seven stars with definitely
underestimated $L_{\rm bol}$.

The stellar radii were estimated by three different methods: using 
$L_{\rm bol}$ and $T_{\rm eff}$ ($R_{\rm bol}$), using the
Kervella–Fouqu\'e ratio ($R_{\rm KF}$), and via $P_{\rm rot}$ and 
$v\sin i$ ($R^{'}_{\rm rot}$). The estimates of $R_{\rm bol}$ and 
$R_{\rm KF}$ were shown to be in excellent agreement. We revealed five 
stars with overestimated $R_{\rm bol}$ ($R_{\rm bol} \gg R\sin i $) and 
two stars with underestimated $R_{\rm bol}$  
($R_{\rm bol}\ll R\sin i$). The possible causes of the discrepancies 
between $R_{\rm bol}$ and $R\sin i $ for these seven stars were discussed.
The estimates of $R_{\rm bol}$ and $R^{'}_{\rm rot}$ for the stars with
known rotation periods $P_{\rm rot}$ and $v\sin i$ were shown
to be in excellent agreement with the mean ratio
$\langle R^{'}_{\rm rot}/R_{\rm bol}\rangle= 1.125 \pm 0.096$. We revealed ten stars
with small inclinations ($\sin i \ll \pi/4$) for which $R^{'}_{\rm rot}$
were definitely underestimated. For the remaining 24
stars from Wichmann’s list with reliable estimates
of $R^{'}_{\rm rot}$, we calculated the individual distances. The
mean distance to these stars was shown to be
$143\pm26$ pc, which is in excellent agreement with the
adopted distance to the Taurus–Auriga SFR.

The stellar masses and ages were determined for
all stars with reliable luminosities. The stars from
Wichmann’s list were found to have masses in the
range from $0.4\,M_\odot$ to $2.2\,M_\odot$ and ages from 1.5 to
100 Myr. About 33\% of the stars from Wichmann’s
list have ages younger than 10 Myr and can be members
of the Taurus–Auriga SFR.

Owing to this work, there is a sample of 74
magnetically active stars toward the Taurus–Auriga
SFR. For all these stars, we determined such basic
physical parameters as the luminosities, radii,
masses, and ages. For 52 objects we know the rotation periods.
Improving the evolutionary status
of these objects and investigating the possible relationship
between the magnetic activity and rotation
of these stars are of indubitable interest. The results
of these studies will be presented in two next papers.

\subsection*{ACKNOWLEDGMENTS}

The observational part of this work was performed
with several telescopes at the Maidanak Astronomical
Observatory of the Ulugh Beg Astronomical Institute
of the Uzbek Academy of Sciences 
during 2003–2006 and was supported by the Center
for Science and Technologies of Uzbekistan (grant
no. F-2-2-3). I wish to thank D. Alekseev, S. Melnikov,
B. Kahharov, O. Ezhkova, and S. Artemenko
for their participation in the photometric observations.

\newpage

\subsection*{REFERENCES}
\noindent
01. M. Ammler, V. Joergens, and R. Neuh\"auser, Astron.
Astrophys. \textbf{440}, 1127 (2005).

\noindent
02. C. Bertout and F. Genova, Astron. Astrophys. \textbf{460},
499 (2006).

\noindent
03. M. Bessell, Astron. J. \textbf{101}, 662 (1991).

\noindent
04. J. Bouvier, R. Wichmann, K. Grankin, et al., Astron.
Astrophys. \textbf{318}, 495 (1997).

\noindent
05. C. Broeg, V. Joergens, M. Fern\'andez, et al., Astron.
Astrophys. \textbf{450}, 1135 (2006).

\noindent
06. C. J. Clarke and J.Bouvier, Mon. Not. R. Astron. Soc.
\textbf{319}, 457 (2000).

\noindent
07. M. Cohen and L. Kuhi, Astrophys. J. Suppl. Ser. \textbf{41},
743 (1979).

\noindent
08. S. A. Ehgamberdiev, A. K. Baijumanov, S. P. Ilyasov,
et al., Astron. Astrophys. Suppl. Ser. \textbf{145}, 293 (2000).

\noindent
09. K. N. Grankin, M. A. Ibragimov, V. B. Kondrat'ev, et
al., Astron. Rep. \textbf{39}, 799 (1995).

\noindent
10. K. N. Grankin, Astron. Lett. \textbf{24}, 497 (1998).

\noindent
11. K. N. Grankin, Astron. Lett. \textbf{25}, 526 (1999).

\noindent
12. K. N. Grankin, S. A. Artemenko, and S. Y. Melnikov,
Inform. Bull. Var. Stars, No. 5752 (2007a).

\noindent
13. K. N. Grankin, S. Yu. Melnikov, J. Bouvier, et al.,
Astron. Astrophys. \textbf{461}, 183 (2007b).

\noindent
14. K. N. Grankin, J. Bouvier, W. Herbst, et al., Astron.
Astrophys. \textbf{479}, 827 (2008).

\noindent
15. P. Hartigan, K. M. Strom, and S. E. Strom, Astrophysics
\textbf{427}, 961 (1994).

\noindent
16. C. de Jager and H. Nieuwenhuijzen, Astron. Astrophys.
\textbf{177}, 217 (1987).

\noindent
17. H. L. Johnson, in \textit{Nebulae and Interstellar Matter},
Ed. by B. M. Middlehurst and L. H. Aller (Univ. of
Chicago Press, Chicago, 1968), p. 167.

\noindent
18. S. J. Kenyon and L. Hartmann, Astrophys. J. Suppl.
Ser. \textbf{101}, 117 (1995).

\noindent
19. P. Kervella and P. Fouqu\'e, Astron. Astrophys. \textbf{491},
855 (2008).

\noindent
20. R. Kohler and C. Leinert, Astron. Astrophys. \textbf{331}, 977
(1998).

\noindent
21. E. L. Martin, in \textit{Cool Stars in Clusters and Associations:
Magnetic Activity and Age Indicators}, Ed.
by G. Micela, R. Pallavicini, and S. Sciortino, Mem.
Soc. Astr. It. \textbf{68}, 905 (1997).

\noindent
22. E. L. Martin and A. Magazz\`u, Astron. Astrophys.
\textbf{342}, 173 (1999).

\noindent
23. A. Massarotti, D.W. Latham, G. Torres, et al., Astron.
J. \textbf{129}, 2294 (2005).

\noindent
24. A. F. L. Nemec and J. M. Nemec, Astron. J. \textbf{90}, 2317
(1985).

\noindent
25. D. C. Nguyen, R. Jayawardhana, M.H. van Kerkwijk,
et al., Astrophysics \textbf{695}, 1648 (2009).

\noindent
26. P. Sartoretti, R. A. Brown, D. W. Latham, et al.,
Astron. Astrophys. \textbf{334}, 592 (1998).

\noindent
27. L. Siess, E. Dufour, and M. Forestini, Astron. Astrophys.
\textbf{358}, 593 (2000).

\noindent
28. R.W. Tanner, JRASC \textbf{42}, 177 (1948).

\noindent
29. A. Tokunaga, \textit{Allen’s Astrophysical Quantities}, 4th
ed., Ed. by A. N. Cox (Springer, New York, 2000),
p. 143.

\noindent
30. G. Torres, J. Andersen, and A. Gim\'enez, Astron.
Astrophys. Rev. \textbf{18}, 67 (2010).

\noindent
31. R.Wichmann, J. Krautter, J.H.M.M. Schmitt, et al.,
Astron. Astrophys. \textbf{312}, 439 (1996).

\noindent
32. R. Wichmann, G. Torres, C. H. F. Melo, et al., Astron.
Astrophys. \textbf{359}, 181 (2000).

\noindent
33. Li-Feng Xing, Xiao-Bin Zhang, and Jian-Yan Wei,
Chin. J. Astron. Astrophys. \textbf{6}, 716 (2006).

\end{document}